\documentclass[a4paper]{article}

\usepackage{itgspeech2025}    
\usepackage{times}            
\usepackage[english]{babel}   
\usepackage[utf8]{inputenc}
\usepackage[T1]{fontenc}      
\usepackage[sort&compress,numbers]{natbib}	
\usepackage{amsmath,amssymb}
\usepackage{graphicx}
\usepackage[colorlinks=false,pdfborder={0 0 0}]{hyperref}
\usepackage{units}
  
\usepackage{microtype}
\usepackage{type1cm}

\usepackage{booktabs}
\usepackage{tabularx}
\usepackage{flushend} 

\title{Comparison of Knowledge Distillation Methods for Low-complexity Multi-microphone Speech Enhancement using the FT-JNF Architecture}

\author{Robert Metzger$^{1,2}$, Mattes Ohlenbusch$^1$, Christian Rollwage$^1$, Simon Doclo$^{1,3}$}

\address{%
\small%
$^1$\,Fraunhofer Institute for Digital Media Technology IDMT, Oldenburg Branch for Hearing, Speech and Audio Technology HSA\\
$^2$\,OFFIS Institute for Information Technology Oldenburg\\
$^3$\,Dept. of Medical Physics and Acoustics and Cluster of Excellence Hearing4all, Carl von Ossietzky Universität Oldenburg\\
Email: \texttt{robert.metzger@offis.de}
}

\begin{document}

\maketitle

\begin{abstract}
Multi-microphone speech enhancement using deep neural networks (DNNs) has significantly progressed in recent years.
However, many proposed DNN-based speech enhancement algorithms cannot be implemented on devices with limited hardware resources.
Only lowering the complexity of such systems by reducing the number of parameters often results in worse performance.
Knowledge Distillation (KD) is a promising approach for reducing DNN model size while preserving performance.
In this paper, we consider the recently proposed Frequency-Time Joint Non-linear Filter (FT-JNF) architecture and investigate several KD methods to train smaller (student) models from a large pre-trained (teacher) model.
Five KD methods are evaluated using direct output matching, the self-similarity of intermediate layers, and fused multi-layer losses. 
Experimental results on a simulated dataset using a compact array with five microphones show that three KD methods substantially improve the performance of student models compared to training without KD.
A student model with only 25\% of the teacher model’s parameters achieves comparable PESQ scores at 0\,dB SNR. 
Furthermore, a reduction of up to 96\% in model size can be achieved with only a minimal decrease in PESQ scores. 
\end{abstract}

\section{Introduction}
Environmental noise reduces speech quality and intelligibility and affects listening efforts in many speech communication applications, such as voice control, hearing aids, and teleconferencing systems.
To address this issue, several algorithms based on deep neural networks (DNNs) have been proposed to reduce environmental background noise and improve speech quality~\cite{haebumbach2024, wang2020complex, mack2020deep, Hu.2020, Fu.2022, wang2023tfgridnet,Tesch.2023, sinha2023low, tammen2025imposing}.
However, the computational power and memory requirements of many DNN-based speech enhancement algorithms are too high to be implemented on resource-constrained devices.
Reducing the size of DNNs is a common strategy to enable deployment on such devices. However, this often comes at the cost of significant performance degradation~\cite{zhang_beyond_2024}.
To address this issue, knowledge distillation (KD) has emerged as a promising solution. KD is a training technique in which a smaller student model learns to mimic the behavior of a larger, high-performing teacher model to reduce model complexity while aiming to maintain performance~\cite{AllenZhu.}. KD can also be used, e.g., to retain part of the performance of a non-causal teacher for training a causal student DNN~\cite{Wakayama.2024}, or for training smaller models computing auxiliary speaker embeddings for personalized enhancement models~\cite{serre2025contrastive}.
Recent work has explored the use of KD in the context of speech enhancement, aiming to improve the efficiency of DNN-based approaches~\cite{Cheng., Han.2024, Nathoo.2024}. In~\cite{Cheng.} it was proposed to leverage multiple intermediate layers of a teacher network for distillation, rather than solely relying on its final outputs, to train a smaller student model.
For this purpose, a method was proposed that enables the use of KD even with layers of different dimensions. A central element of this method is the computation of a self-similarity matrix during training. Similarly, in~\cite{Han.2024} it was proposed a cross-layer similarity method between different model architectures to reduce the size of a deep complex convolutional recurrent neural network by 30\% without compromising its performance. In~\cite{Nathoo.2024} several ways of computing self-similarity between teacher and student models were investigated, and a two-step training strategy was proposed that improved the student model performance.

In this paper, we investigate KD for a multi-microphone speech enhancement system using the Frequency-Time Joint Non-linear Filter (FT-JNF) architecture~\cite{Tesch.2023}. A teacher model with 1400k parameters serves as the performance reference. We evaluate five KD methods using direct output matching and self-similarity between outputs of intermediate layers to train student models with reduced complexity. 
The student models trained via KD minimized a soft loss function that measured the difference between intermediate representations from various student layers and those of a pre-trained, larger teacher model.
In the first experiment, different KD methods are investigated for training a student model with a size of 44k parameters and compared to a model trained without KD. 
In the second experiment, the impact of KD on performance at different model sizes varying from 13k to 365k parameters is investigated. All experiments are conducted on simulated five-microphone data using reverberant speech and directional noise, and performance is evaluated using PESQ scores across multiple signal-to-noise ratios (SNRs). Experimental results show that the size of the student FT-JNF can be reduced by up to 96\% compared to the teacher model while only minimally affecting the PESQ score.

\section{Signal model}
We consider a noisy, reverberant acoustic environment with a single talker being recorded by a microphone array with $M$ microphones.  
In the short-time Fourier transform (STFT) domain, the reverberant speech component $X_m(k,l) \in \mathbb{C}$ of the $m$-th microphone at frame index $l$ and frequency index $k$ can be modeled\footnote{This approximation is valid if the STFT frame is longer than the acoustic impulse response.} as the product of the clean anechoic speech STFT coefficient $S(k, l)$ and the acoustic transfer function $H_m(k)$ between the talker and the $m$-th microphone:
\begin{equation}
    X_m(k,l) = S(k,l) \cdot H_m(k).
    \label{eq:mixture-signal}
\end{equation}
The noisy microphone signal consists of the reverberant speech component and the noise component $V_m(k,l)$ record\-ed by the $m$-th microphone, i.e.  
\begin{equation}
    Y_m(k,l) = X_m(k,l) + V_m(k,l).
    \label{eq:noise-speech-add}
\end{equation}
%
\section{Multi-channel speech enhancement system}

\begin{figure*}[ht]
    \centering
    \includegraphics[width=\linewidth]{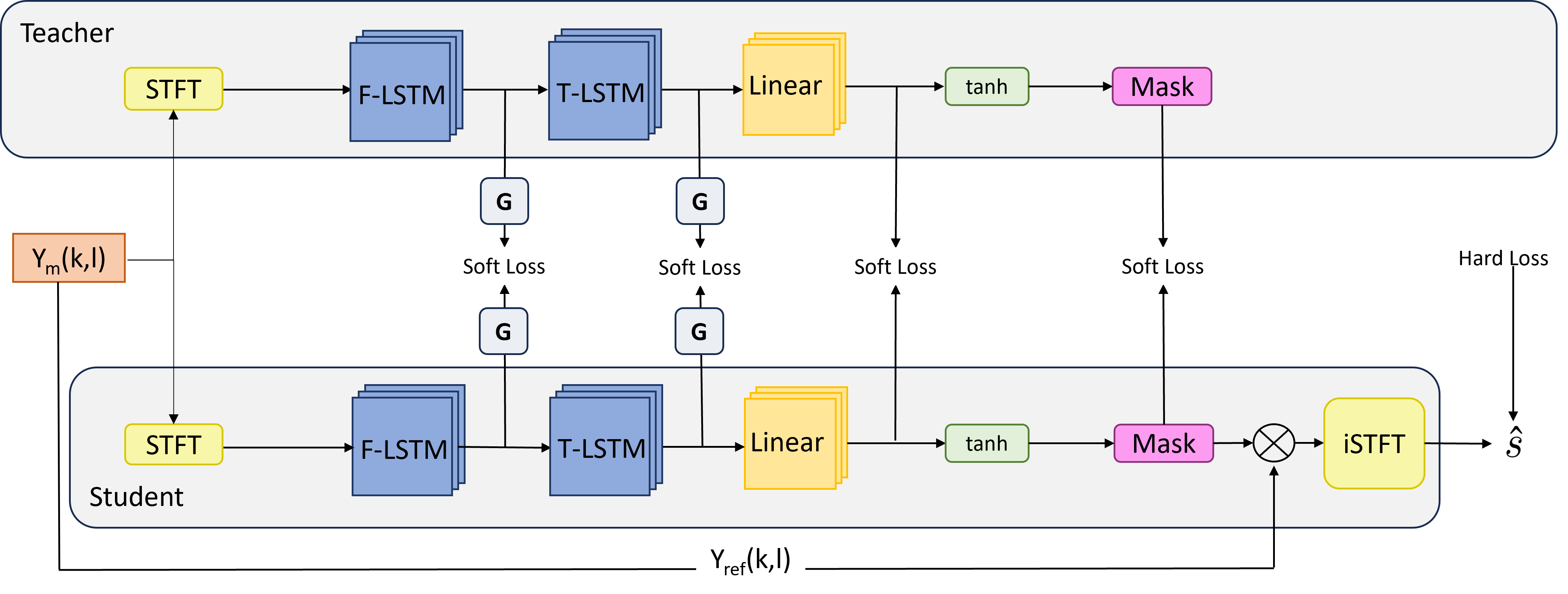}
    \caption{Schematic overview of KD techniques with the FT-JNF architecture, depicting the teacher model (top) and the student model (bottom). The hard loss is exclusively computed from the output of the student model. The soft loss is either computed directly from the activation outputs of both models, which have identical shape, or from self-similarity matrices $\mathbf{G}$ constructed from intermediate LSTM outputs, which differ in shape.
    }
    \label{fig:KD-Methods}
\end{figure*}

In this paper, we consider the multi-microphone speech enhancement system proposed in \cite{Tesch.2023}, which is based on the FT-JNF architecture (see Fig. \ref{fig:KD-Methods}).
It takes the real and imaginary parts of noisy STFT coefficients of all microphones as input. 
The architecture consists of a long short-term memory (LSTM) layer operating across frequency (F-LSTM), an LSTM layer operating across time (T-LSTM), a linear layer and a $\tanh$ activation function. Different from \cite{Tesch.2023}, no decompression was applied after the $\tanh$ activation.
The outputs are the real and imaginary parts of a single-channel complex-valued mask $W(k,l)$ which is applied to the noisy reference microphone signal $Y_\text{ref}(k,l)$, i.e.
\begin{equation}
    \hat{S}(k,l) = W(k,l)\cdot Y_\text{ref}(k,l).
    \label{eq:filter-beamformen}
\end{equation}
In this paper, we will consider different model sizes by changing the number of hidden units in the LSTM layers (see Table~\ref{tab:model-sizes}).
\begin{table}[ht]
    \centering
    \caption{Considered model sizes with corresponding numbers of hidden units in the LSTM layers, number of parameters, number of MACs per frame, and file size in MB. Size A corresponds to the teacher model; size E corresponds to the model investigated in Section~\ref{sec:results_methods}.}
    \small
    \begin{tabular}{ccrrr}
    \toprule
         & \textbf{F/T-LSTM} & \textbf{Param}\,[$10^3$] & \textbf{MACs}\,[$10^9$] & \textbf{Size}\,[MB] \\
     \midrule
         \textbf{A}  & \textbf{512 / 256} & \textbf{1400} & \textbf{34.7} & \textbf{9.64}\\
         B & 256 / 64 & 364.9 & 8.9 & 3.49 \\
         C & 128 / 32 & 92.7 & 2.3 & 1.44 \\
         D & 88 / 40 & 56.4 & 1.4 & 1.1 \\
         \textbf{E} & \textbf{80 / 32} & \textbf{44.4} & \textbf{1.1} & \textbf{0.95} \\
         F & 72 / 24 & 33.9 & 0.85 & 0.65 \\
         G & 64 / 16 & 24.9 & 0.63 & 0.64 \\
         H & 56 / 8 & 17.4 & 0.44 & 0.55 \\
         I & 48 / 8 & 13.4 & 0.34 & 0.48 \\
     \bottomrule
    \end{tabular}
    \label{tab:model-sizes}
\end{table}

\section{Knowledge Distillation Methods}
\label{sec:KD-Methods}
KD~\cite{Hinton.} is a technique to train DNNs, where a (large) teacher model is used to train a smaller student model, aiming at achieving better performance than training the student model without the teacher.
As the teacher, the FT-JNF model with size A is trained using the combined $L_1$-loss in time and STFT domain (after re-analysis)~\cite{wang_stft-domain_2023} between the clean anechoic speech signal $s(n)$ and the output signal $\hat{s}(n)$, i.e.
\begin{equation}
\begin{split}
    L_\text{hard}\left(\hat{s}(n), s(n)\right) = \sum_n\lvert \hat{s}(n) - &s(n)\rvert \\  + \sum_{k,l} \bigl|\lvert  \text{STFT}\{\hat{s}(n)\}(k,l)\rvert-  &\lvert\text{STFT}\{s(n)\}(k,l)\rvert\bigr|,
\end{split}
\label{eq:hardloss}
\end{equation}
where $n$ denotes the discrete time index. This loss function is referred to as hard loss.
To train the student models, we consider different KD methods by combining the hard loss in~\eqref{eq:hardloss} on the output of the student model with soft losses $L_\text{soft}$ computed using intermediate outputs of the teacher model.
As intermediate outputs we will consider the model output, the output of the linear layer, or the output of the LSTM layers (see Sections~\ref{sec:KD-nofusion} and~\ref{sec:KD-selfsim}).
For all KD methods, we follow the two-stage training approach proposed in~\cite{Nathoo.2024}. The loss function for this approach is given by
\begin{equation}
    L = \alpha \cdot L_\text{hard} + (1- \alpha) \cdot L_\text{soft},
    \label{eq:Loss}
\end{equation}
where $\alpha$ represents a weighting factor.
In the first stage, the student model is pre-trained to minimize the soft loss up to the selected layers of the student and the teacher models, i.e. using $\alpha=0$.
In the second stage, the student model is further trained by minimizing the hard loss, i.e. using $\alpha=1$.
\subsection{KD without fusion method}
\label{sec:KD-nofusion}

For some KD methods, the soft loss can be directly computed as
\begin{equation}
    L_\text{soft}\left(\mathbf{Z}^\text{teacher}, \mathbf{Z}^\text{student}\right) =  \lVert \mathbf{Z}^\text{teacher} - \mathbf{Z}^\text{student}\rVert_1,
    \label{eq:l1-loss} 
\end{equation}
where $\mathbf{Z}^\text{teacher}$ and $\mathbf{Z}^\text{student}$ denote the intermediate output of the teacher and the student, respectively, and $\lVert ... \rVert_1$ denotes the $L_1$-Norm.
Since this method requires both intermediate outputs to have the same shape, this soft loss can only be applied to the masks at the model output (KD Mask) and the output of the linear layer (KD Linear) of FT-JNF models with different model sizes.
\subsection{KD with self-similarity matrix}
\label{sec:KD-selfsim}
 
Unlike the linear layers and the masks, the intermediate outputs of the LSTM layers have different shapes, i.e. model sizes, between the teacher and the student.
Since it is not possible to compute an element-wise soft loss, as in~\eqref{eq:l1-loss}, it was proposed in~\cite{Nathoo.2024} to compute a soft loss using self-similarity matrices instead.
Consider the outputs of intermediate layers $\mathbf{Z}^\text{teacher} \in \mathbb{R}^{(K\cdot L)\times C^\text{teacher}}$ and $\mathbf{Z}^\text{student} \in \mathbb{R}^{(K\cdot L)\times C^\text{student}}$, where $K$ and $L$ denote the number of frequency bins and frames, and $C^\text{teacher}$ and $C^\text{student}$ denote the number of hidden units for the considered layers of the teacher and student models (with $C^\text{student}<C^\text{teacher}$). The self-similarity matrices for teacher and student $\mathbf{G}^\text{teacher} \in \mathbb{R}^{(K\cdot L)\times(K\cdot L)}$ and $\mathbf{G}^\text{student} \in \mathbb{R}^{(K\cdot L)\times(K\cdot L)}$ can then be computed as the Gram matrices
$\mathbf{G}^\text{teacher}=\mathbf{Z}^\text{teacher} \left(\mathbf{Z}^\text{teacher}\right)^{T}$ and 
$\mathbf{G}^\text{student}=\mathbf{Z}^\text{student} \left(\mathbf{Z}^\text{student}\right)^{T}$,
where $(\cdot)^T$ denotes the transpose. The soft loss can then be defined as $L_1$-loss between the self-similarity matrices, i.e.
\begin{equation}
    L_\text{soft}\left(\mathbf{Z}^\text{teacher}, \mathbf{Z}^\text{student}\right) =  \lVert \mathbf{G}^\text{teacher} - \mathbf{G}^\text{student}\rVert_1.
\end{equation}
Since this method does not rely on teacher and student models having the same intermediate output shapes, it can be used to perform KD on the output of the F-LSTM layer (KD~F-LSTM) and the T-LSTM layer (KD~T-LSTM). 
In addition, we consider a KD method where the soft losses for both LSTM layers and the linear layer are added (KD Multi).
Following \cite{AllenZhu.}, we expect that KD Multi will provide the best results, as most information is exchanged between the teacher and student models.
Table \ref{tab:KD-methods} presents an overview of all considered KD methods and their use of intermediate layer outputs and fusion methods.

\begin{table}
    \centering
    \caption{KD methods considered for training and the layers and fusion methods used.}
    \hspace{0.1cm}
    \small
    \begin{tabularx}{\columnwidth}{ccc}
    \toprule
         \textbf{KD Method} & \textbf{Used Layers} & \textbf{Fusion method} \\
     \midrule
         KD Mask & Masks & - \\
         KD Linear & Linear & - \\
         KD F-LSTM & F-LSTM & self-similarity \\
         KD T-LSTM & T-LSTM & self-similarity \\
         KD Multi & F-LSTM, T-LSTM, Linear & self-similarity \\
     \bottomrule
    \end{tabularx}
    \label{tab:KD-methods}
\end{table}

\section{Experimental setup}
 To investigate the effectiveness of the considered KD methods in training the FT-JNF speech enhancement system, an experimental evaluation was conducted for a compact microphone array with five microphones at different model sizes and complexities.

\subsection{Datasets and performance metrics}
All simulations were conducted at a sampling frequency of 16\,kHz, using an STFT framework with a frame length of 32\,ms, a frame shift of 16\,ms, and a square-root Hann analysis and synthesis window.
A part of the German split of the Common Voice dataset (v11.0)~\cite{Mozilla.2022} was used as clean speech dataset, 
split into disjunct subsets of 38 hours for training, 4.8 hours for validation, and 1.2 hours for testing.
The noise signals from the ICASSP 2023 Deep Noise Suppression Challenge~\cite{Dubey.2024} were used as noise dataset for training and evaluation. 
Room impulse responses (RIRs) from the open source collection openSLR SLR26~\cite{Ko.2017} were used to simulate reverberation. 

\begin{figure}
    \centering
    \begin{minipage}[b]{0.49\columnwidth}
        \centering
        \includegraphics[width=\linewidth]{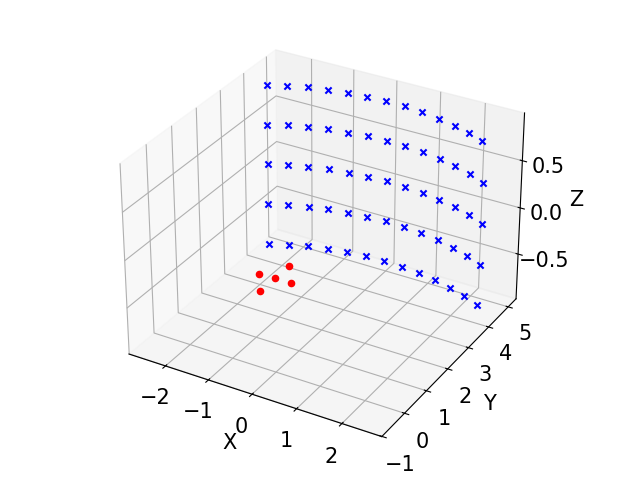}
    \end{minipage}
    \hfill
    \begin{minipage}[b]{0.49\columnwidth}
        \centering
        \includegraphics[width=\linewidth]{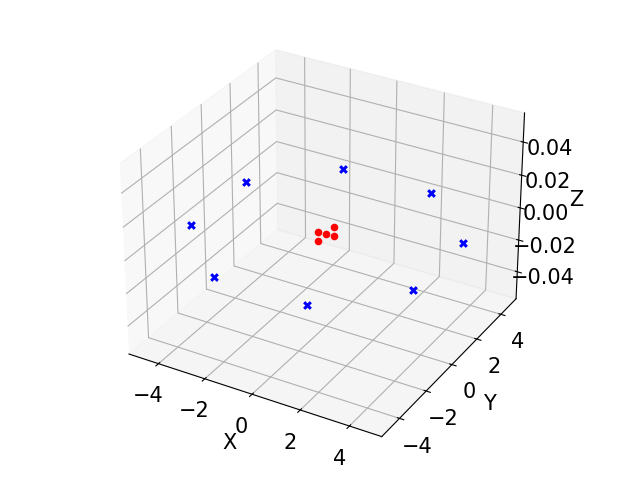}
    \end{minipage}
    \caption{Simulated positions of the target talker (left) and the noise sources (right) used in training and evaluation. }
    \label{fig:source-position}
\end{figure}
As shown in Figure \ref{fig:source-position}, the target talker positions were simulated 5\,m from the microphone array at an azimuth of $\pm 30^\circ$ (steps of $5^\circ$) and an elevation of $\pm10^\circ$ (steps of $5^\circ$), assuming free field propagation between the target talker and the microphones. This scenario was used in order to achieve a preferred direction for the speech enhancement system.
For the background noise, eight noise source positions were simulated at an elevation of $0^\circ$ around the microphone (azimuth steps $45^\circ$).  
For each example during training as well as evaluation, a random target talker position and noise source position were selected, and randomly selected speech and noise signals were convolved with time-delayed pulses between the source and the microphones and a randomly selected RIR. The simulated speech and noise components at the microphones were mixed at SNRs between -5 and 15\,dB defined at the front microphone for training and evaluation. The performance of the models was evaluated using the wideband PESQ score \cite{international_telecommunications_union_itu_itu-t_2001} of the output signal, using the clean anechoic speech signal $s(n)$ as the reference signal.
The center microphone of the array was used as the reference microphone in~\eqref{eq:filter-beamformen}. 

\subsection{Training procedure}
\label{sec:training_details}
All models were trained using the Adam optimizer~\cite{kingma_adam_2015} with an initial learning rate of $5\cdot10^{-4}$.
The learning rate was halved whenever the validation loss did not improve for three consecutive epochs.
The teacher model was trained for a maximum of 100 epochs, and early stopping was used if the validation loss did not improve for six consecutive epochs.
For the KD-based training of the student models, the two-stage approach from~\cite{Nathoo.2024} was employed by first training for up to 100 epochs using $\alpha=0$ in~\eqref{eq:Loss}, with the same early stopping criterion used as for training the teacher model. In the second stage, a maximum of 100 epochs and the same early stopping criterion were employed using $\alpha=1$ in~\eqref{eq:Loss}. The learning rate was reset to its original value at the start of the second training stage.
Each training step employed audio examples with 4 seconds length and a batch size of 4.

\section{Results}
This section presents the results of the experimental evaluation. In Section~\ref{sec:results_methods}, the performance of different KD methods is compared across different SNRs for a fixed student model size. In Section~\ref{sec:results_scaling}, the performance for different student model sizes is investigated for a fixed SNR.

\subsection{KD methods}
\label{sec:results_methods}
\begin{figure}
    \centering
        \includegraphics[width=\linewidth]{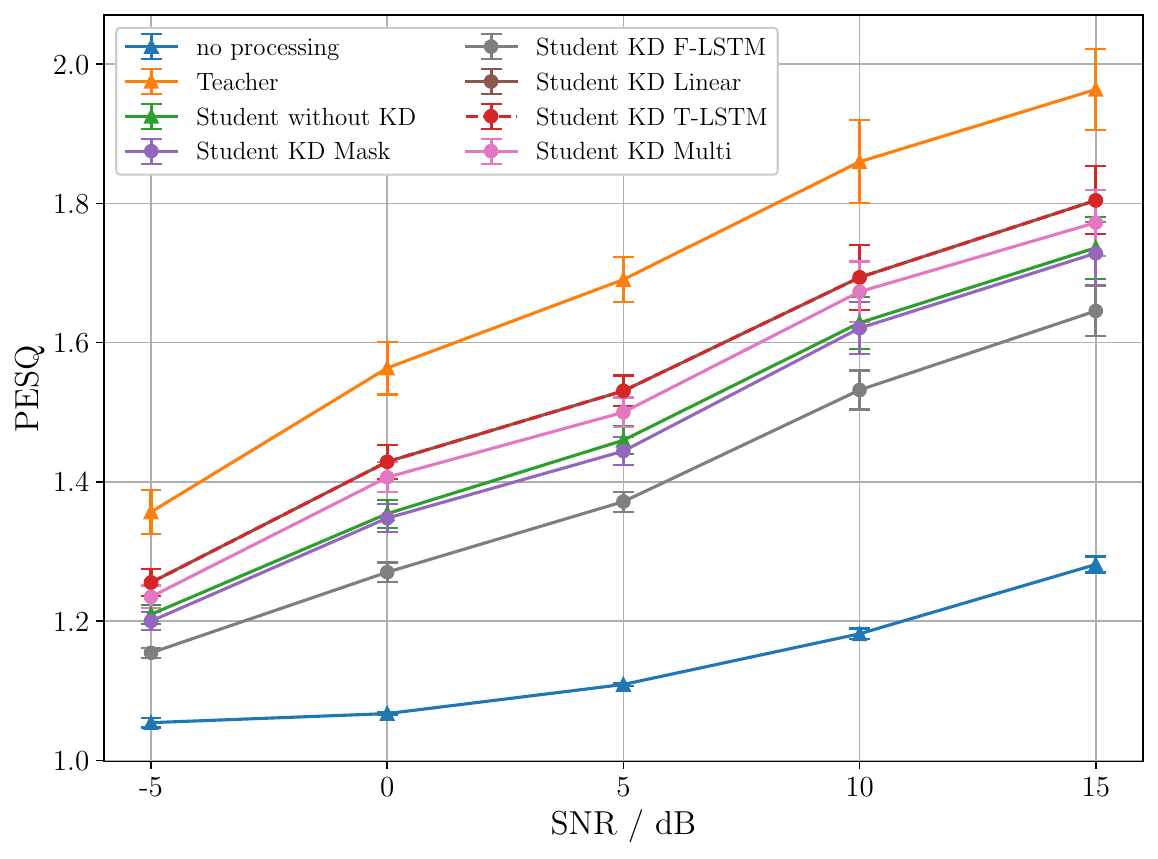}
        \caption{Median PESQ scores and variance for different SNRs for models trained using different KD methods (see Table \ref{tab:KD-methods}). Results are compared to the teacher model (size A), a student model trained without KD, and the unprocessed reference microphone signal.}
        \label{fig:method-comparison}
\end{figure}

Figure \ref{fig:method-comparison} presents the median PESQ scores and their variance of various model configurations for SNRs ranging from -5 to 15\,dB. The blue curve corresponds to the unprocessed noisy reference microphone signal, while the orange curve corresponds to the teacher model. The other curves correspond to student models with model size E (see Table \ref{tab:model-sizes}), either trained without KD or trained with the KD methods outlined in Section \ref{sec:KD-Methods}.
\begin{figure}
        \centering
        \includegraphics[width=\linewidth]{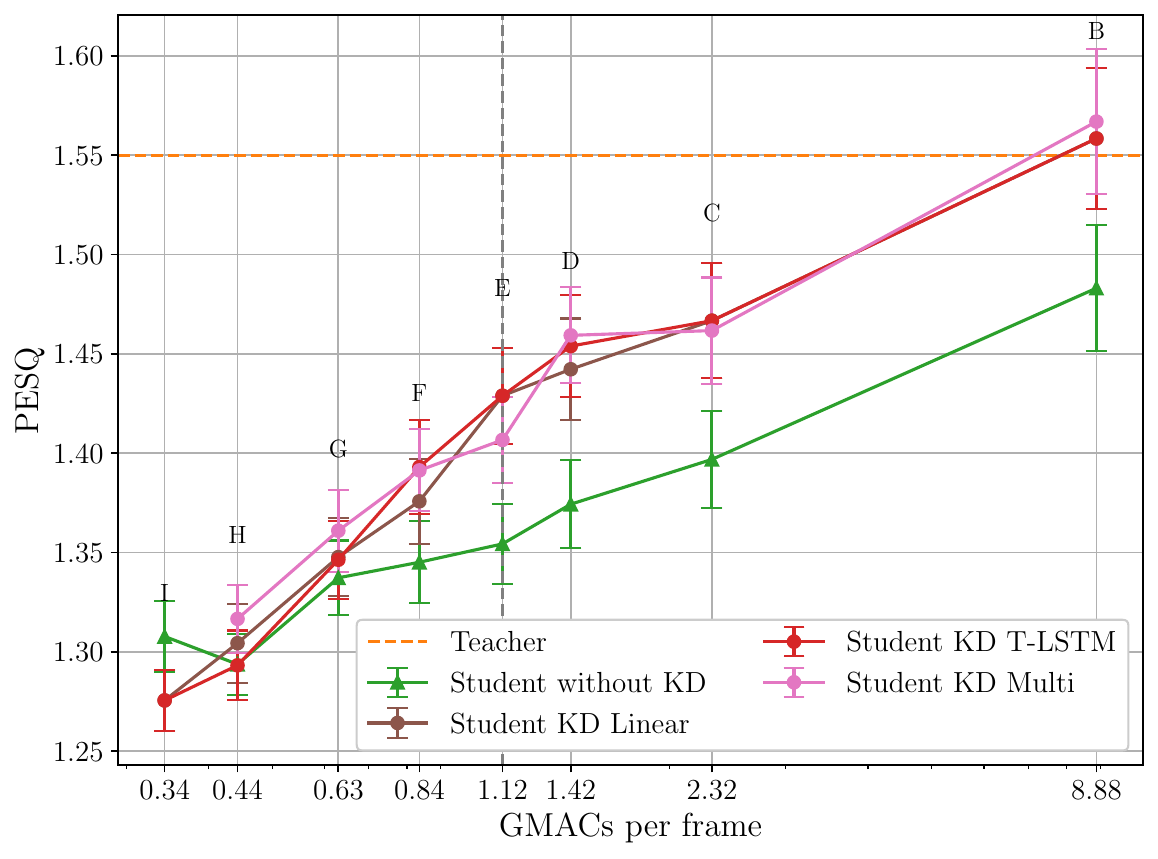}
        \caption{Median PESQ scores and variance as a function of GMACs per frame for the teacher model (size A) and for student models of varying sizes, either trained without KD or using KD~Linear, KD~Multi, and KD~T-LSTM. The SNR was equal to 0\,dB. The gray dashed line marks model size E used in the previous experiment.}
        \label{fig:size-comparesion}
\end{figure}%
First, it can be observed that the PESQ scores of all models increase with SNR and that the (large) teacher model outperforms all (smaller) student models.
Among the student models, the use of KD during training results in higher PESQ scores for three of the considered KD methods (namely KD T-LSTM, KD~Linear and KD Multi) compared to the student model trained without KD. Student KD Linear and T-LSTM yield nearly identical results, with minor differences despite visual overlap.
The student model trained with KD~Mask yielded a similar PESQ score to the student model trained without KD, presumably because the soft loss on the mask is very similar to the hard loss on the output signal. The student model trained with KD F-LSTM even degraded performance compared with the student model trained without KD, indicating that KD only improves performance when using layers closer towards the output of the model.
The following section considers only the KD methods that improve performance over training the student model without KD (KD~Linear, KD~T-LSTM, KD~Multi).

\subsection{Model scaling}
\label{sec:results_scaling}

For an SNR of 0\,dB, Figure \ref{fig:size-comparesion} presents the PESQ score of student models with various model sizes (B to I, see Table~\ref{tab:model-sizes}), either trained without KD or trained with KD Linear, KD T-LSTM or KD Multi, as a function of MACs per frame.
Model size E used in the previous section corresponds to the gray dashed line.
The yellow dashed line corresponds to the performance of the teacher model (size A).
An upward trend in PESQ scores is observed across all models as the model size increases. 
From model size F onwards, all KD-trained student models significantly outperform the student model trained without KD, where only minor differences can be observed among the considered KD methods. 
KD-trained student models of size B (25\% of the teacher model size A) achieve a PESQ score comparable to the PESQ score of the teacher model. KD-trained student models with a size below F show no advantage over the student model trained without KD.

\section{Conclusion}
In this paper, we investigated the use of KD for training a multi-microphone speech enhancement system based on the FT-JNF architecture with low computational complexity.
The KD-trained student models minimized a soft loss between intermediate outputs of different layers and those of a pre-trained, larger teacher model.
Experimental results showed that student models trained with KD methods using layers closer towards the output of the model in the FT-JNF architecture significantly increased performance relative to the student model trained without KD. 
Furthermore, results showed that KD-trained student models are able to achieve a comparable PESQ score as the teacher model with a four times smaller model size.
If a reduction of 0.1 in the PESQ score relative to the teacher model is considered acceptable, a model size reduction of 96\% can be achieved by training a student model using KD.
\newpage
\subsection*{Acknowledgment}
The Oldenburg Branch for Hearing, Speech and Audio Technology HSA is funded in the program \frqq Vorab\flqq~by the Lower Saxony Ministry of Science and Culture (MWK) and the Volkswagen Foundation for its further development.
This work was partly funded by the Deutsche Forschungsgemeinschaft (DFG, German Research Foundation) - Project ID 352015383 - SFB 1330 C1.

\small
\bibliographystyle{ieeetr}
\bibliography{ITG_manual_bib}


\end{document}